\documentclass[3p,twocolumn]{elsarticle}

\usepackage{verbatim} 
\usepackage{threeparttable}

\begin{document}

\title{Monte Carlo modeling of spallation targets containing uranium and americium}

\author[fias]{Yury Malyshkin}
\ead{malyshkin@fias.uni-frankfurt.de}
\author[fias,inr]{Igor Pshenichnov}
\ead{pshenich@fias.uni-frankfurt.de}
\author[fias,kurch]{Igor Mishustin}
\author[fias]{Walter Greiner}

\address[fias]{Frankfurt Institute for Advanced Studies, J.-W. Goethe University, 60438 Frankfurt am Main, Germany}
\address[inr]{Institute for Nuclear Research, Russian Academy of Sciences, 117312 Moscow, Russia}
\address[kurch]{National Research Center ``Kurchatov Institute'', 123182 Moscow, Russia}

\begin{abstract}

Neutron production and transport in spallation targets made of uranium and
americium are studied with a Geant4-based code MCADS (Monte Carlo model for
Accelerator Driven Systems). A good agreement of MCADS results with experimental data on neutron-
and proton-induced reactions on $^{241}$Am and $^{243}$Am nuclei allows to use
this model for simulations with extended Am targets.
It was demonstrated that MCADS model can be used for 
calculating the values of critical mass for $^{233,235}$U, 
$^{237}$Np, $^{239}$Pu and $^{241}$Am.
Several geometry options and material compositions (U, U+Am, Am, Am$_2$O$_3$)
are considered for spallation targets to be used in 
Accelerator Driven Systems. All considered options operate as deep subcritical targets having neutron
multiplication factor of $k\sim 0.5$. It is found that more than 4~kg of Am can be burned
in one spallation target during the first year of operation.

\end{abstract}

\begin{keyword}
spallation reactions \sep minor actinides \sep neutron sources \sep Accelerator Driven Systems \sep radioactive waste
\end{keyword}

\maketitle

\newpage

\section{Introduction}\label{sec:intro}

Many neutrons can be produced in spallation nuclear reactions~\cite{Bauer2001,Filges2010} induced by
energetic protons in collisions with heavy target nuclei like W, Ta, Bi and Pb due to their enhanced neutron content with
respect to lighter nuclei. This method to create an intense flux of neutrons
is known for decades and it is already employed in several existing~\cite{SNS,ISIS} spallation neutron sources
and will be used in the facilities to be constructed, e.g., in the ESS project~\cite{ESS}.  Such facilities are
dedicated to neutron imaging and scattering experiments~\cite{Arai2009}.  Accelerator Driven Systems (ADS) aimed at energy
production in subcritical assemblies of fissile materials or burning nuclear
waste~\cite{Herrera-Martinez2007a,Herrera-Martinez2007} also use an intense proton beam to produce neutrons 
in spallation targets. The design of a spallation target is a challenging part of such projects in view of high 
energy deposited by the proton beam and secondary particles and the radiation damage of the target material. 
The performance of a target irradiated by a megawatt-power proton beam was the subject of a 
dedicated experiment~\cite{Wagner2008}.

Heavy materials like W, Ta, Bi and Pb are commonly used in the design of spallation targets. Although the fission of such
nuclei can, in principle, be induced by energetic protons~\cite{Borger2013}, its role in neutron production is 
negligible. However, an alternative approach can be also considered to involve fissionable, 
$^{232}$Th, $^{238}$U~\cite{Hashemi-Nezhad2011}, or even fissile,
$^{235}$U, $^{242\rm m}$Am~\cite{Galy2002}, materials in the design of a spallation target. The difference between these
two groups of materials consists in the capability of fissile material to sustain a nuclear chain reaction once
a critical mass of this material is accumulated. Such materials can be either directly
irradiated by a proton beam, or used as a blanket surrounding a non-fissionable material impacted by protons.
In both cases neutron production is boosted due to additional fission neutrons.
As recently demonstrated by our calculations~\cite{Malyshkin2012a}, the number of neutrons produced per beam proton is
about 3 times higher in a uranium target compared to one made of tungsten, while the energy deposition
calculated per produced neutron remains comparable in both targets. Therefore, a less powerful beam is needed to
achieve the same neutron flux in the uranium target as in the tungsten target, and the total energy deposition
in both targets~\cite{Malyshkin2012a} remain comparable. Thermal energy released in fission reactions can be 
converted to electricity and then support, at least in part, the operation of the accelerator.

Apart from the need to build intense neutron sources, using fissile materials in spallation targets opens the
possibility to transmute them in fission reactions induced by primary protons and secondary nucleons.
Indeed, in addition to unused uranium, each 1000~kg of spent nuclear fuel discharged
from a light-water reactor typically contain several kilograms of fissile transuranium elements like plutonium
and Minor Actinides (MA): neptunium, americium and curium~\cite{Salvatores2011}. Up to 99.9\% of plutonium can be
extracted and then further used in nuclear reactors~\cite{Rome1991}. However, other radioactive elements,
MA and long-lived fission products, are still very hazardous due to their high radiotoxicity, and
their release to environment has to be avoided. There are plans to confine them in very robust vitrified blocks
stored in deep geological repositories. Alternatively, MA contained in spent nuclear fuel can be separated 
and recycled in a dedicated facility operating with fast neutrons (as thermal neutrons are not efficient).  
As demonstrated by many dedicated studies, see e.g.~\cite{Salvatores2011}, the extracted MA can be
efficiently transmuted into short-lived or stable fission products in fast reactors or in accelerator-driven reactor cores.  

Certainly, more theoretical and experimental studies are needed to design an intense fast-neutron source or a 
spallation target containing fissionable or fissile materials.
For many years experimental studies of transmutation of
long-lived radiotoxic nuclides have been carried out at the Joint Institute for Nuclear Research in Dubna, Russia,
in the framework of an international collaboration~\cite{Westmeier2011}. In particular, $^{237}$Np and $^{241}$Am
were transmuted into short-lived or stable nuclides by neutrons produced by protons in a thick lead target.
Within the project called ``Energy plus Transmutation'' beams of protons and deutrons were used, and the flux of
fast neutrons was amplified by a massive uranium sleeve surrounding a non-fissile
target~\cite{Borger2013,Borger2011,Borger2012}.

Detailed theoretical modeling of ADS prototypes should precede their construction and operation.
Therefore, a reliable computational tool based on modern software is necessary to foster
studies in the field of the accelerator-driven transmutation.
A number of Monte Carlo codes have been used to simulate neutron production and 
transport in spallation targets of ADS: PHITS~\cite{Sato2013_PHITS}, 
SHIELD~\cite{Batyaev2008_SHIELD}, MCNPX~\cite{McKinney2012_MCNPX} and others. 
However, to the best of our knowledge, spallation targets containing Am were 
not studied with these codes so far.
In the present work we further develop our
Geant4-based code MCADS (Monte Carlo model for Accelerator Driven Systems)~\cite{Malyshkin2012a,Malyshkin2012} in order to apply 
it for fissile spallation targets containing U and Am. Modeling spallation targets containing americium is motivated 
by the following two reasons~\cite{Salvatores2011}. First, americium is the most
abundant MA in spent nuclear fuel and its transmutation into relatively short-lived fission products can reduce 
the radiotoxicity of radioactive waste by an order of magnitude. Second, the operation of fast reactors with a high 
content of MA causes certain safety concerns.  Alternatively, a subcritical system driven by an accelerator could be a 
promising option to burn americium extracted from spent nuclear fuel.

\section{Modeling of americium transmutation by slow and energetic nucleons}\label{sec:validation}

As demonstrated in our previous works~\cite{Malyshkin2012a,Malyshkin2012}, 
all physics processes relevant to neutron generation and transport in conventional non-fissile and also in
fissionable uranium targets can be successfully simulated
with the Geant4 toolkit~\cite{Agostinelli2003,Allison2006, Apostolakis2009}. In particular, these 
processes include  spallation and fission reactions induced by primary protons and secondary nucleons. 
Usually specific Geant4 simulations are performed with a set of physical models 
known as a Physics List. 

All present calculations were performed with Geant4 of version 9.4 with 
patch 01 as in our previous works [13,22]. In this version of the toolkit
the following models are available for simulating $p$-nucleus interactions:
Bertini Cascade, Binary Cascade and Intra-Nuclear Cascade Li\`ege
coupled with the fission-evaporation model ABLA. These models 
are included in the QGSP\_BERT\_HP, QGSP\_BIC\_HP and 
QGSP\_INCL\_ABLA Physics Lists, respectively. 
The prefix QGSP indicates that quark-gluon string model is used for high-energy
interactions. All three Physics Lists employ High Precision (HP) 
model for neutron interactions below 20~MeV which use
evaluated nuclear data libraries described below. The ionization energy loss 
of charged particles was simulated with Standard Electromagnetic Physics 
package of Geant4. The physics models used in Geant4 are described in detail
in Geant4 Physics Manual~\cite{G4PhysManual}.

In Ref.~\cite{Malyshkin2012a} we have evaluated the performance of the 
above-mentioned physics models for tungsten and uranium targets irradiated 
by protons. Fission cross sections and multiplicities of 
neutrons produced in thin uranium targets by protons with energies 
of 27, 63 and 1000~MeV were calculated and compared with experimental 
data~\cite{Isaev2008, Bernas2003}. It was demonstrated, that the 
INCL\_ABLA~\cite{Boudard2002,Heikkinen2008} better describes the data as 
compared with other models. In particular, only INCL\_ABLA predicts the fission 
cross section and the neutron multiplicity for 1000~MeV protons very close to 
data, within the uncertainty of the measurements. However, one can note that 
all the considered cascade models become less accurate for proton energies below 
100~MeV~\cite{Malyshkin2012a}. 

The average numbers of neutrons produced in extended tungsten and uranium targets 
irradiated by 400-1500 MeV protons were also calculated and compared with experimental data,
see Ref.~\cite{Malyshkin2012a}. As shown, also in this case the combination of 
INCL\_ABLA and NeutronHP models provides the most accurate results, which differ by less than 
10\% from the experimental data. The Bertini Cascade model mostly overestimates, 
while the Binary Cascade model underestimates the neutron yields. 
Therefore, we conclude that the QGSP\_INCL\_ABLA\_HP Physics List is the best choice 
among other options for simulating nuclear reactions in uranium and tungsten 
targets. The aforementioned lack of accuracy for nucleons with energies below 
100~MeV does not affect significantly the results, as such
nucleons do not dominate in the considered spallation targets.

In order to perform simulations with materials containing americium several extensions of the Geant4 toolkit 
have been introduced in~\cite{Malyshkin2013a}.
This made possible the simulations of proton- and neutron-induced nuclear reactions and elastic scattering of nucleons on 
Am and other transuranium nuclei. 
In our recent publication~\cite{Malyshkin2013} the (p,f), (n,f) and (n,$\gamma$) cross
sections as well as mass distributions
of fission fragments, average number of neutrons per fission event and secondary neutron spectra were calculated with
MCADS for $^{241}$Am and $^{243}$Am, and good agreement with experimental data was obtained. This justifies 
using MCADS to simulate extended targets containing $^{241}$Am and $^{243}$Am.

\begin{figure}[htb]
\begin{centering}
\includegraphics[width=1.0\columnwidth]{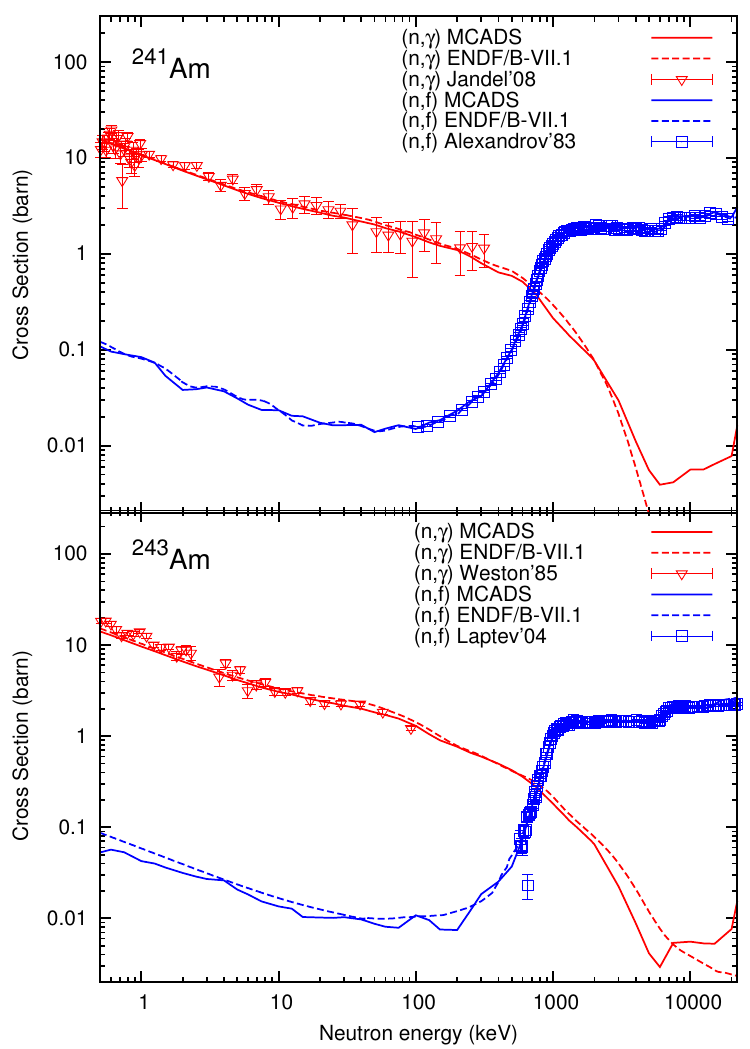}
\caption{Radiative neutron capture cross section (n,$\gamma$), shown in red, and neutron-induced fission
cross section (n,f), shown in blue, for $^{241}$Am (top) and $^{243}$Am (bottom) nuclei.
MCADS results are represented by solid lines, cross sections from ENDF/B-VII.1 evaluated nuclear data library -- by dashed lines.
Measured cross sections, (n,$\gamma$) for $^{241}$Am and $^{243}$Am from Refs.~\cite{Jandel2008,Weston1985}
are shown by triangles, and (n,f) cross section from Refs.~\cite{Alexandrov1983,Laptev2004} -- by squares.}
\label{fig:ncapture_vs_fission}
\end{centering}
\end{figure}

The physics of transmutation of $^{241}$Am and $^{243}$Am nuclei by neutron irradiation can be well understood from
Fig.~\ref{fig:ncapture_vs_fission}. 
Depending on the neutron energy both nuclei can either undergo fission or be transformed via (n,$\gamma$) reaction into A+1 isotopes $^{242}$Am and $^{244}$Am.  After $\beta^-$-decay with half-life times 16~h and 10~h these nuclei change finally into long-lived $^{242}$Cm and $^{244}$Cm, respectively.
A sharp rise of the fission cross section at incident neutron energy of $\sim 0.6$~MeV
leads to the dominance of fission of $^{241}$Am and $^{243}$Am over the neutron capture above 1~MeV. Therefore, fast 
neutrons produced in primary spallation reactions and subsequent
neutron-induced fission reactions can be used to burn $^{241}$Am and $^{243}$Am very efficiently.

The radiative neutron capture (n,$\gamma$) and fission (n,f) cross sections calculated by MCADS by means of the Monte Carlo 
modeling of neutron interactions with a thin layer of $^{241}$Am or $^{243}$Am are plotted in
Fig.~\ref{fig:ncapture_vs_fission} together with the corresponding experimental
data~\cite{Jandel2008,Weston1985,Alexandrov1983,Laptev2004}. Nuclear reactions induced by neutrons with energy
below 20~MeV are simulated by MCADS on the basis of the evaluated nuclear data library JENDL-4.0~\cite{JENDL4}
converted into a format readable by Geant4~\cite{Mendoza2012}. It was found that the Geant4-compatible nuclear data files based on JENDL-4.0 provide the most accurate description of the energy spectra of secondary neutrons 
with respect to other nuclear data libraries. 
The MCADS results below 20~MeV can be compared to the cross sections
extracted from ENDF/B-VII.1~\cite{Chadwick2011}, which are also shown in Fig.~\ref{fig:ncapture_vs_fission}. As seen from this figure,
a very good agreement is obtained between MCADS, experimental data and ENDF/B-VII.1 data.

\section{MCADS calculations of neutron multiplication in fissile materials}\label{sec:criticality}

The key issue in designing a spallation target containing fissile materials is the calculation of the neutron
multiplication factor to ensure that the target operates in a safe subcritical regime.
Neutron multiplication factor $k$ is calculated with MCADS as the ratio between the numbers of
neutrons in the present and previous generations of neutrons averaged over many simulated events. 
The obvious condition is to keep $k<1$, i.e. strictly in the
subcritical mode. The number of neutrons in the target is determined by the balance between their production,
absorption and leakage through the target surface. 

In order to validate the MCADS model with respect to generation of fission neutrons and their absorption
in (n,$\gamma$) and (n,f) processes, we have performed simulations of neutron multiplication in bare (unreflected) 
spheres made of several fissile materials listed in Table~\ref{table:criticality}. 
The radius of each sphere made of specific material was gradually increased until $k$ asymptotically exceeded 1, 
and the mass of such a sphere was defined as the 
critical mass for the given material, see Table~\ref{table:criticality}.
The critical mass data published by the European Nuclear Society~\cite{ENS} and Monte Carlo simulation results
obtained with JENDL-3.2 library 
for $^{237}$Np and $^{241}$Am in Refs.~\cite{YudinNonproliferation} and~\cite{Dias2003} are also presented in Table~\ref{table:criticality}. The results of calculations with MCADS
for $^{233,235}$U, $^{237}$Np and $^{239}$Pu agree within 2--8\% with the published data~\cite{ENS}, but diverge
by $\sim 16$\% for $^{241}$Am. We attribute this deviations to uncertainties of the nuclear data for $^{241}$Am.
As discussed in Ref.~\cite{YudinNonproliferation}, the calculated critical mass of
$^{241}$Am sphere is very sensitive to the cross sections of neutron-induced reactions tabulated in evaluated
nuclear data libraries. The critical mass varies from 55 to 106~kg depending on the nuclear data library used in
simulations, while the results obtained with the same library (JENDL-3.2), but with different codes
(Polina~\cite{YudinNonproliferation} and MCNP~\cite{Dias2003}) agree quite well,
see Table~\ref{table:criticality}. New measurements
and new evaluations of nuclear data for $^{241}$Am are required to reduce these discrepancies. 
Presently, the lowest estimate of the critical mass of $^{241}$Am ($\sim 55$~kg) should be considered as a
conservative safety limit. Much higher values of the critical mass are reported for $^{243}$Am: from 155.7
to 548.6~kg~\cite{YudinNonproliferation} and from 143 to 284~kg~\cite{Dias2003}, again depending on the data
library used in simulations. This indicates the degree of uncertainties in nuclear data available for $^{243}$Am.

\begin{table*}[!htb]
\caption{Calculated critical masses (kilograms) of bare spheres made of $^{233,235}$U,
$^{237}$Np, $^{239}$Pu and $^{241}$Am. Data from European Nuclear Society~\cite{ENS} and Monte Carlo modeling
results by Polina~\cite{YudinNonproliferation} and MCNP~\cite{Dias2003} codes both based on JENDL-3.2 library are given
for comparison.}
\centering
\begin{tabular}{lcccc}
\noalign{\smallskip}
\hline\noalign{\smallskip}
Material   &  MCADS results  &  data~\cite{ENS} & calculations~\cite{YudinNonproliferation} & calculations~\cite{Dias2003} \\
\hline\noalign{\smallskip}
$^{233}$U  &  16.1         &  15.8 &       &      \\
$^{235}$U  &  48.4         &  46.7 &       &      \\
$^{237}$Np &  62.4         &  63.6 &  75.0 &      \\
$^{239}$Pu &  10.8         &  10.0 &       &      \\
$^{241}$Am &  66.7         &  57.6 &  71.8 & 73.7 \\
\hline\noalign{\smallskip}
\end{tabular}
\label{table:criticality}
\end{table*}

\begin{figure}[htb]
\begin{centering}
\includegraphics[width=1.1\columnwidth]{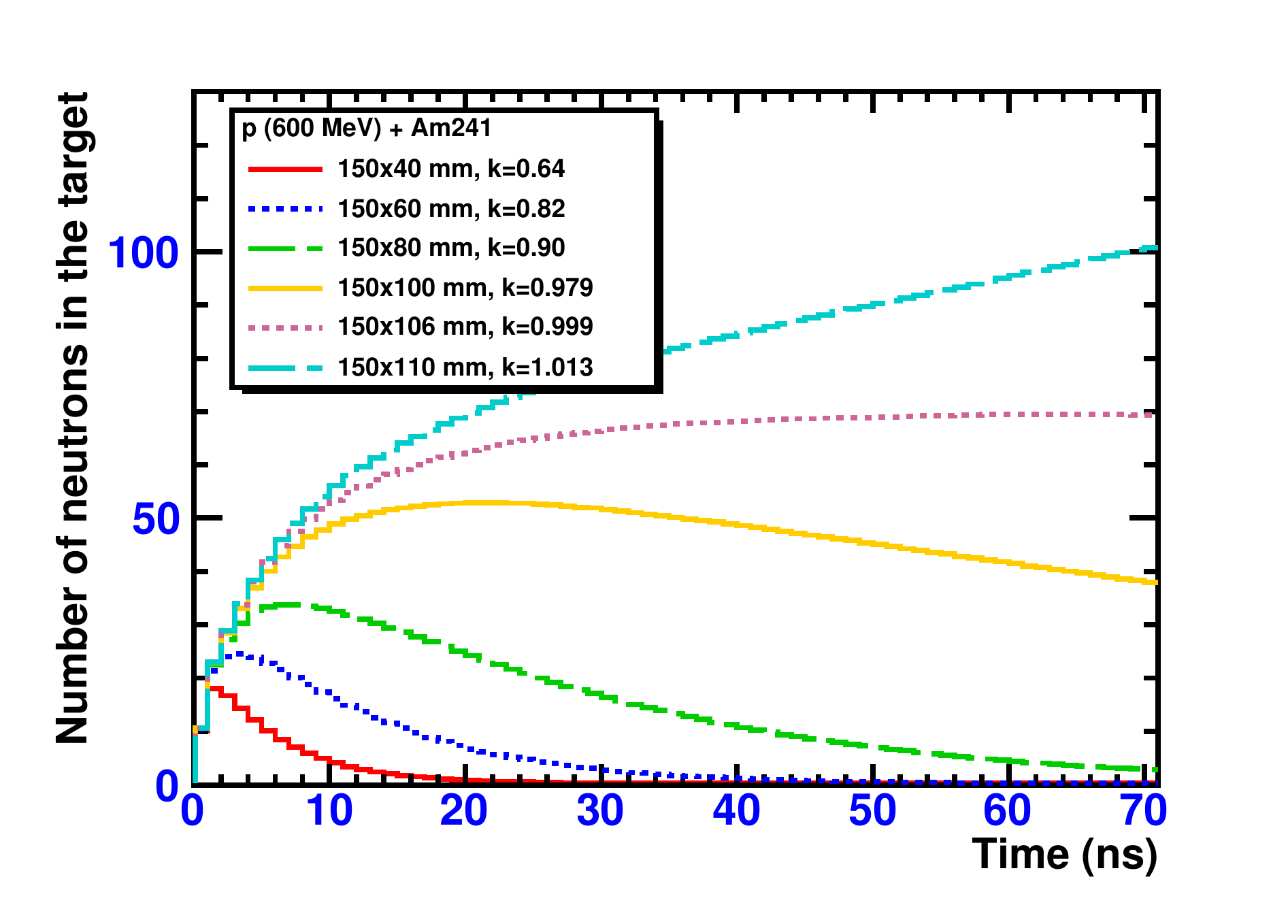}
\caption{Number of neutrons as a function of time in the cylindrical targets made of $^{241}$Am with
the length of 150~mm and various radii and irradiated by 600~MeV protons. The neutron number is normalized
per beam particle. The steady-state behavior (horizontal line) corresponds to approaching critical regime with 
$k=0.999$.}
\label{fig:n_in_target}
\end{centering}
\end{figure}

Following the validation of MCADS results for the critical mass of the $^{241}$Am sphere without external
irradiation, we investigated the criticality issues for a cylindrical spallation target made of pure $^{241}$Am
and irradiated by a proton beam. The length of the target was fixed at 150~mm to ensure that all beam
protons are stopped in the target material, while the target radii were varied from 40~mm to 110~mm.
It was assumed that the target was irradiated by a 600~MeV proton beam with the transverse beam profile of
20~mm FWHM. In Fig.~\ref{fig:n_in_target} we show the time dependence of the average number of neutrons inside the targets
with the target radii of 40, 60, 80, 100, 106 and 110~mm. One can see that the average number of 
neutrons in the targets with radii 40--100~mm  decreases at late time because less neutrons are produced inside 
the target volume than escape it or lost in nuclear interactions. The number of neutrons saturates in the target of 106~mm radius (with the weight of 72.4~kg) just 30~ns after the impact of a beam proton. This case is very close to the critical regime with $k=0.999$. 
Even a smaller fraction of neutrons escape a thicker target of 110~mm radius, and this target becomes supercritical ($k=1.013$). 
The corresponding neutron multiplication factors are listed in the legend
of Fig.~\ref{fig:n_in_target}.

From this criticality study we can conclude that thin cylindrical $^{241}$Am targets 
with typical radii of $\sim 50$~mm and length of 150~mm irradiated by 600~MeV protons will operate in a
deep subcritical regime with  $k\sim 0.7$. This suggests that an equivalent amount of $^{241}$Am can be used
to build a fissile spallation target. All geometrical configurations of spallation targets containing
$^{241}$Am considered below are also designed to operate in a deep-subcritical mode with $k < 0.6$.

\section{Comparison of targets containing uranium and americium}\label{am:simple}

The operation of spallation targets containing uranium can be simulated with confidence, as the nuclear data for
$^{\rm nat}$U and all its most abundant isotopes are reliable in all versions of nuclear data libraries.
The simulations of pure uranium targets are very instructive for further comparison with 
U+Am and pure Am targets. Moreover, the neutrons from U fission 
can be used for the Am transmutation. Therefore, we have performed simulations of cylindrical targets
made of $^{\rm nat}$U, pure $^{241}$Am, a mixture of $^{241}$Am and $^{243}$Am  (57\% and 43\%, respectively) and
americium oxide Am$_2$O$_3$ with the same isotopic composition of Am. 
The choice of Am$_2$O$_3$ is motivated by
the fact that americium is usually extracted from spent nuclear fuel in the form of americium oxide.
Each target has the radius of 40~mm and
length of 120~mm. All these targets have masses well below the critical mass given in Sec.~\ref{sec:criticality}.
It was assumed that the targets were irradiated by the proton beam
with the FWHM of 20~mm and the energy of 600~MeV. 

\begin{figure*}[!htb]
\begin{centering}
\includegraphics[height=0.23\textheight]{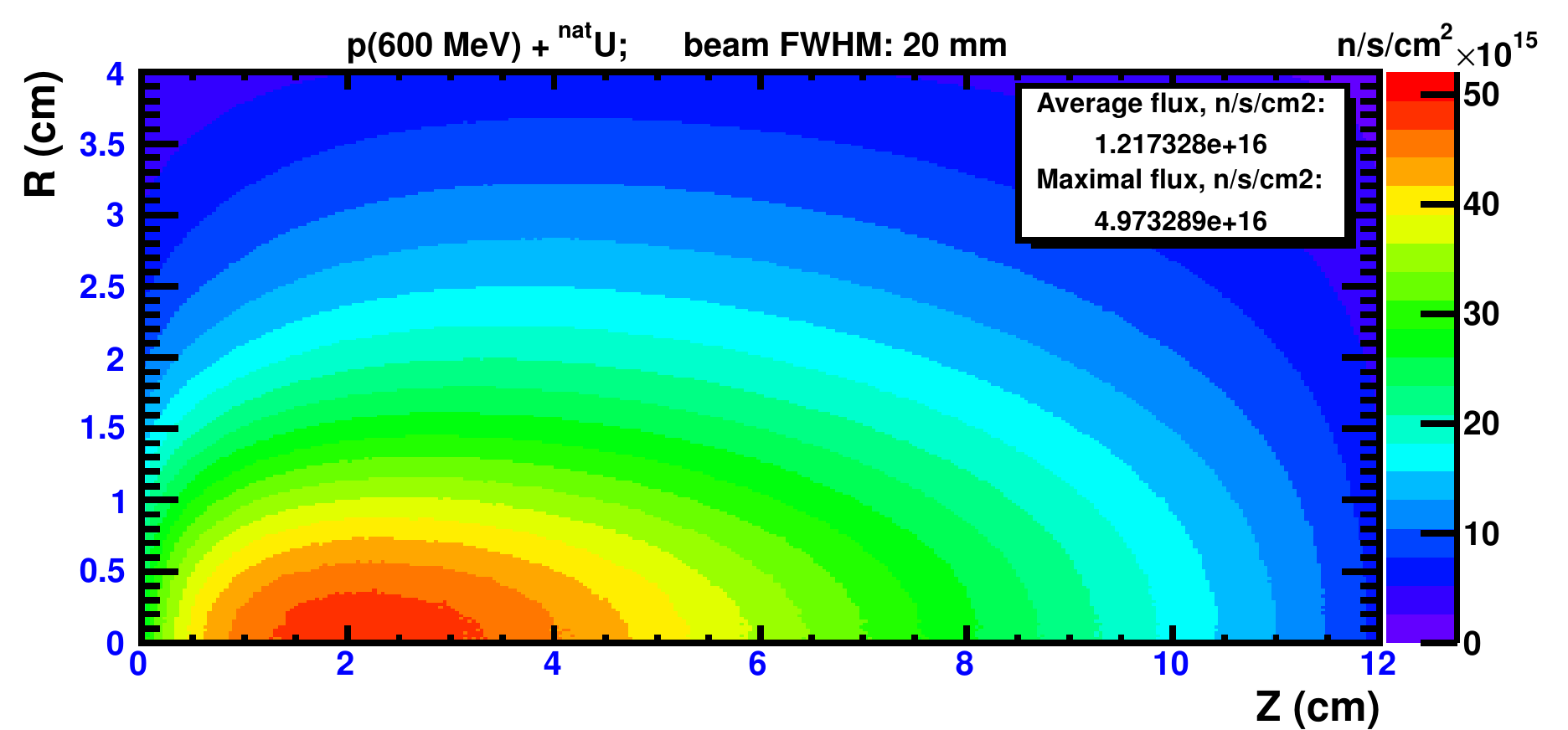}
\includegraphics[height=0.23\textheight]{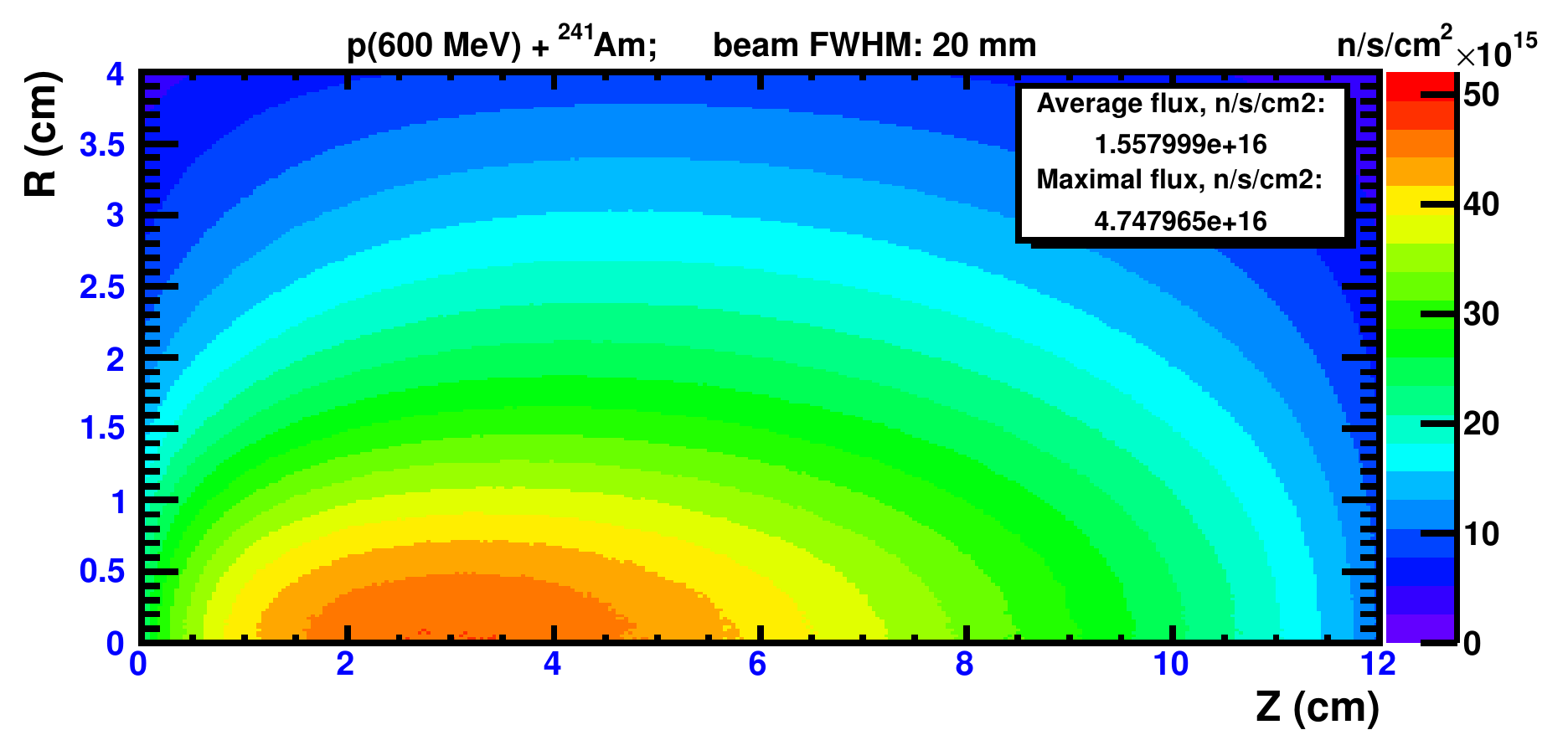}
\includegraphics[height=0.23\textheight]{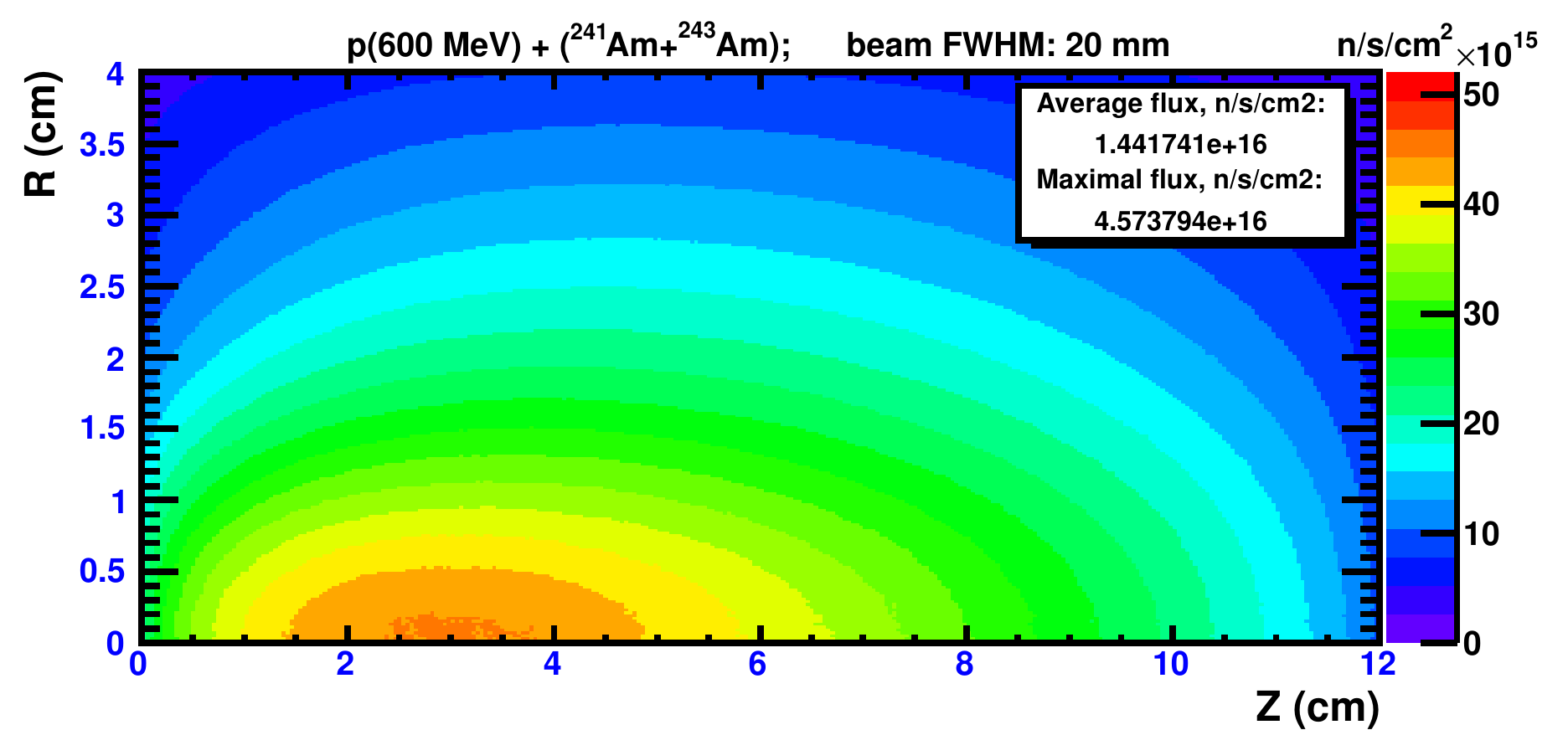}
\includegraphics[height=0.23\textheight]{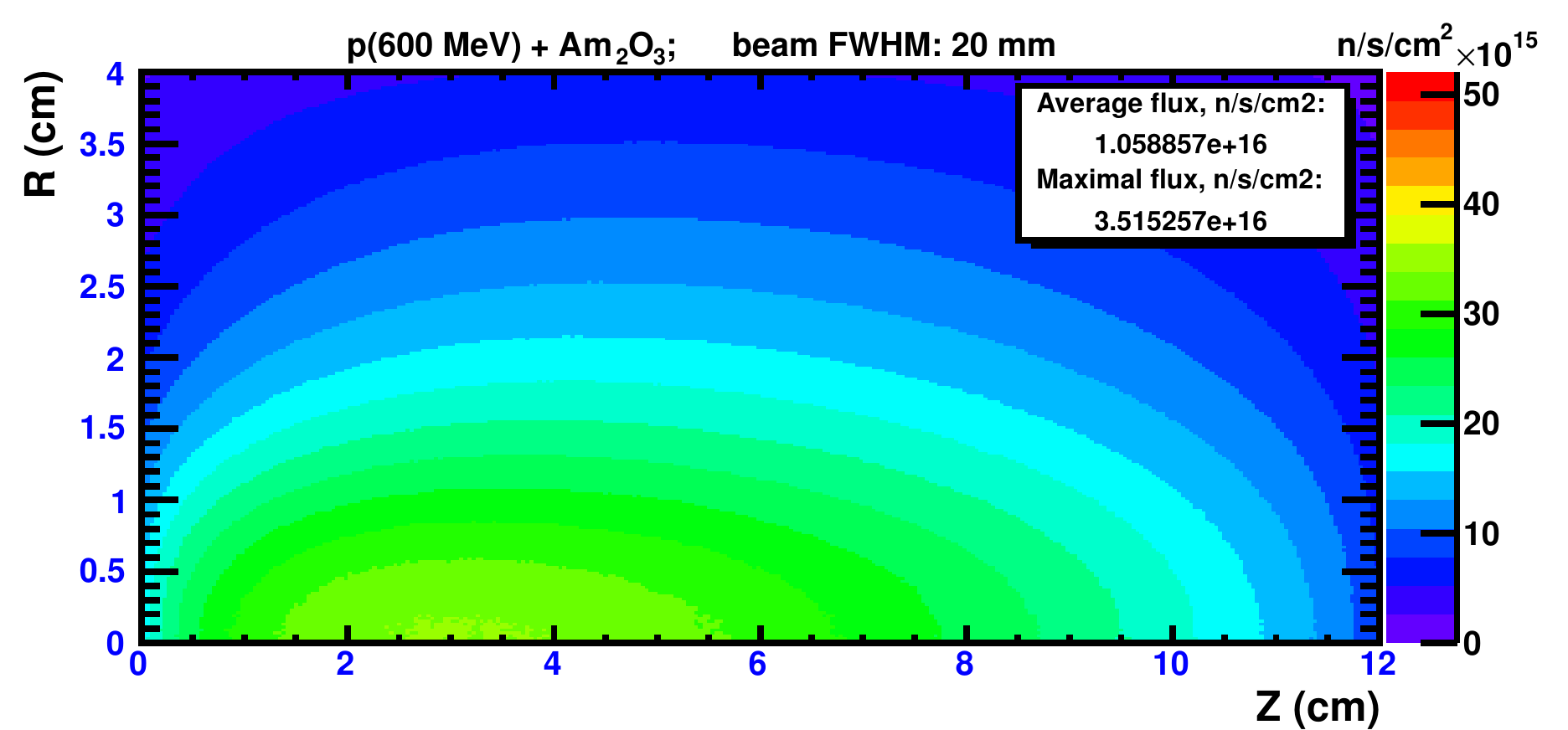}
\caption{Distribution of the neutron flux inside cylindrical targets made of
$^{\rm nat}$U, $^{241}$Am, mixture of $^{241}$Am and $^{243}$Am and Am$_2$O$_3$, all of the same dimensions
specified in Table~\ref{table:Am_vs_U}. The targets are irradiated by a 10~mA 600~MeV proton beam.}
\label{fig:rod_nCF}
\end{centering}
\end{figure*}

The spatial distributions of neutron flux calculated with MCADS for the considered four targets are shown in
Fig.~\ref{fig:rod_nCF}. Although the results are given for the proton current of 10~mA, they can be easily 
rescaled to the actual beam current. The average neutron flux in the americium target ($1.56\cdot10^{16}$~n/s/cm$^2$) is higher
than in the uranium one ($1.22\cdot10^{16}$~n/s/cm$^2$) due to a higher fission
cross section for Am. Since the fission cross section on $^{241}$Am is almost 3 times higher than on U, one could 
expect even a larger difference in favor of $^{241}$Am. However, other reactions, like (n,2n), (n,3n), (n,4n), are 
much more probable on uranium nuclei. As the result, the difference in the average neutron
flux between U and Am targets is reduced. As one can see from Fig.~\ref{fig:rod_nCF}, the results for pure $^{241}$Am
and mixed $^{241}$Am+$^{243}$Am targets are very similar.

\begin{figure*}[!htb]
\begin{centering}
\includegraphics[height=0.23\textheight]{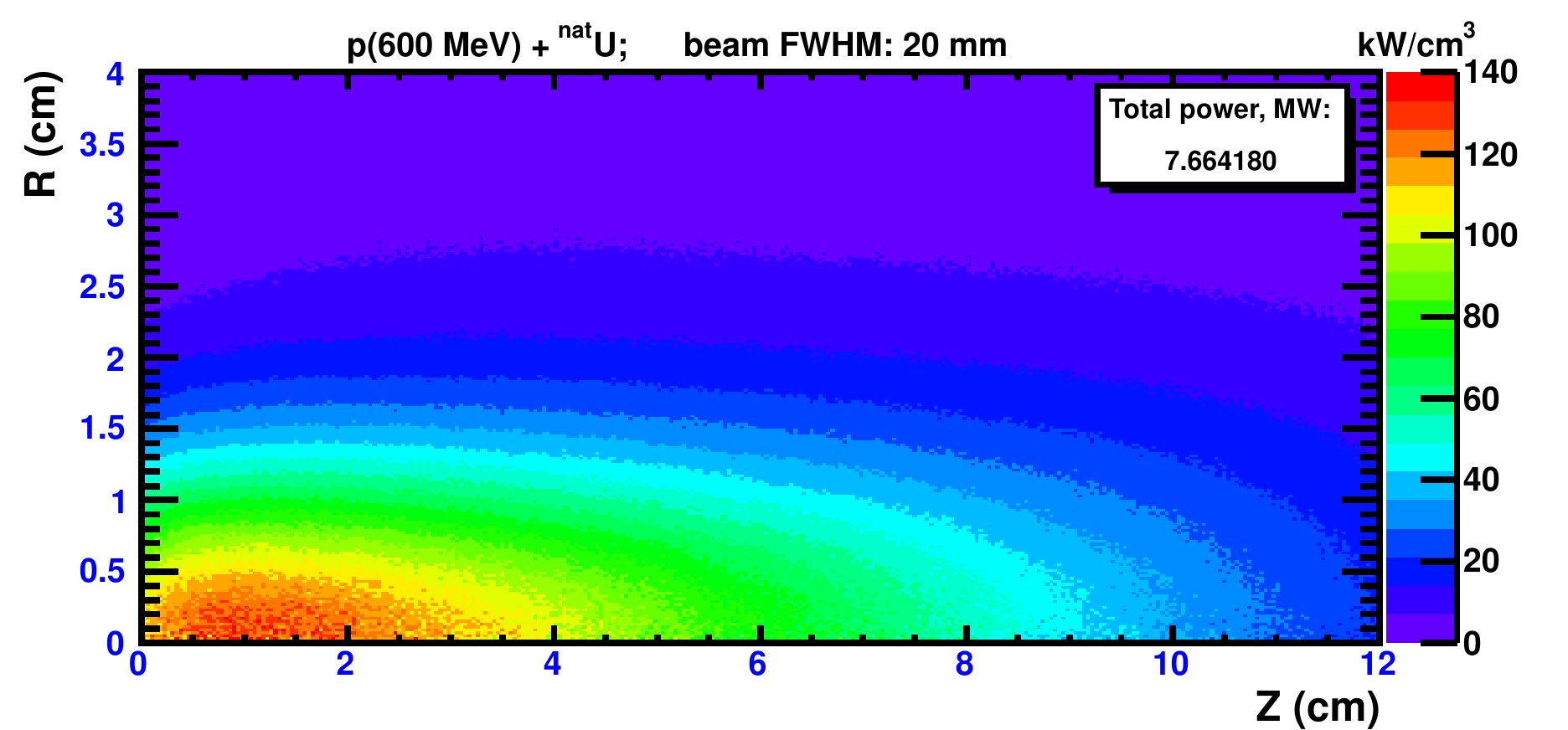}
\includegraphics[height=0.23\textheight]{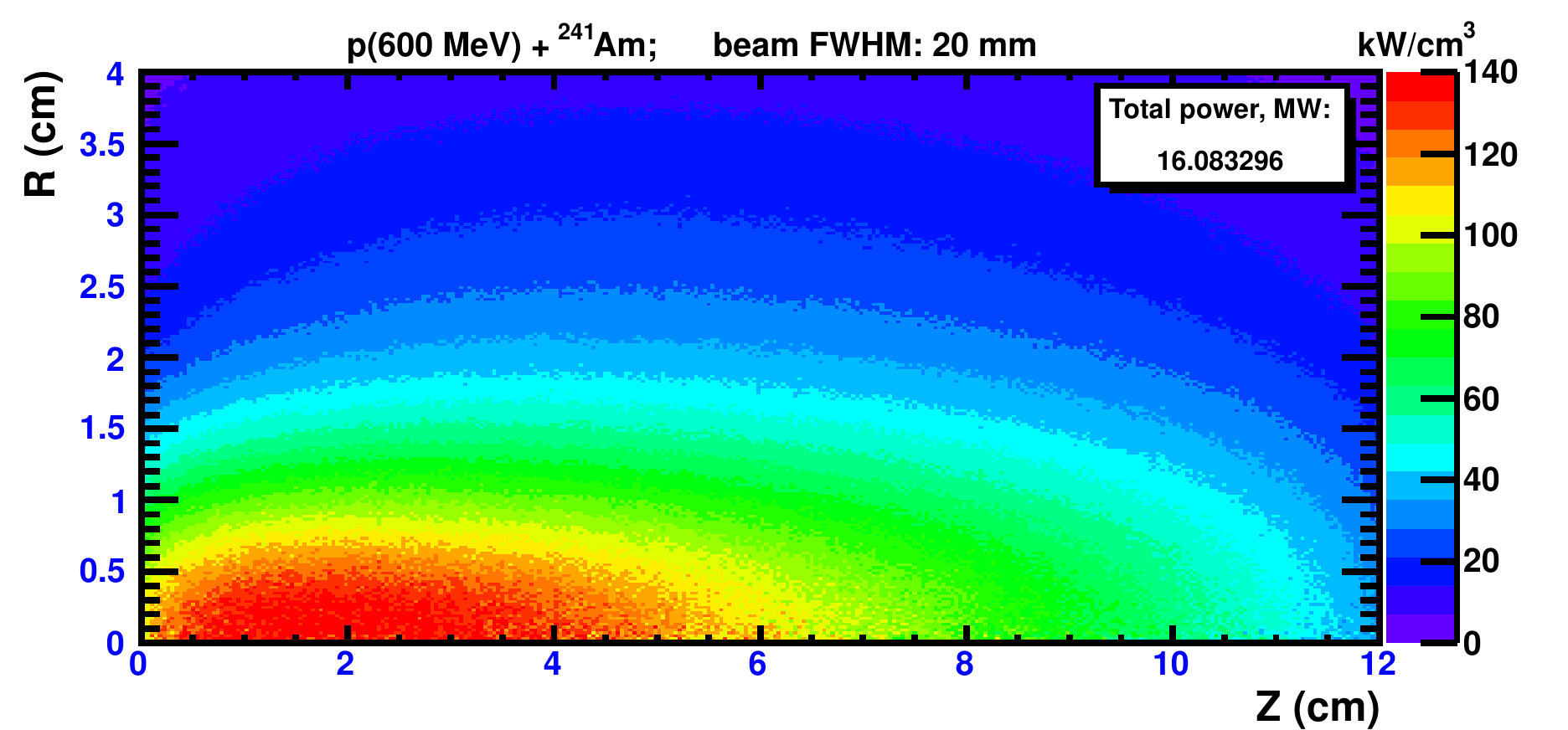}
\includegraphics[height=0.23\textheight]{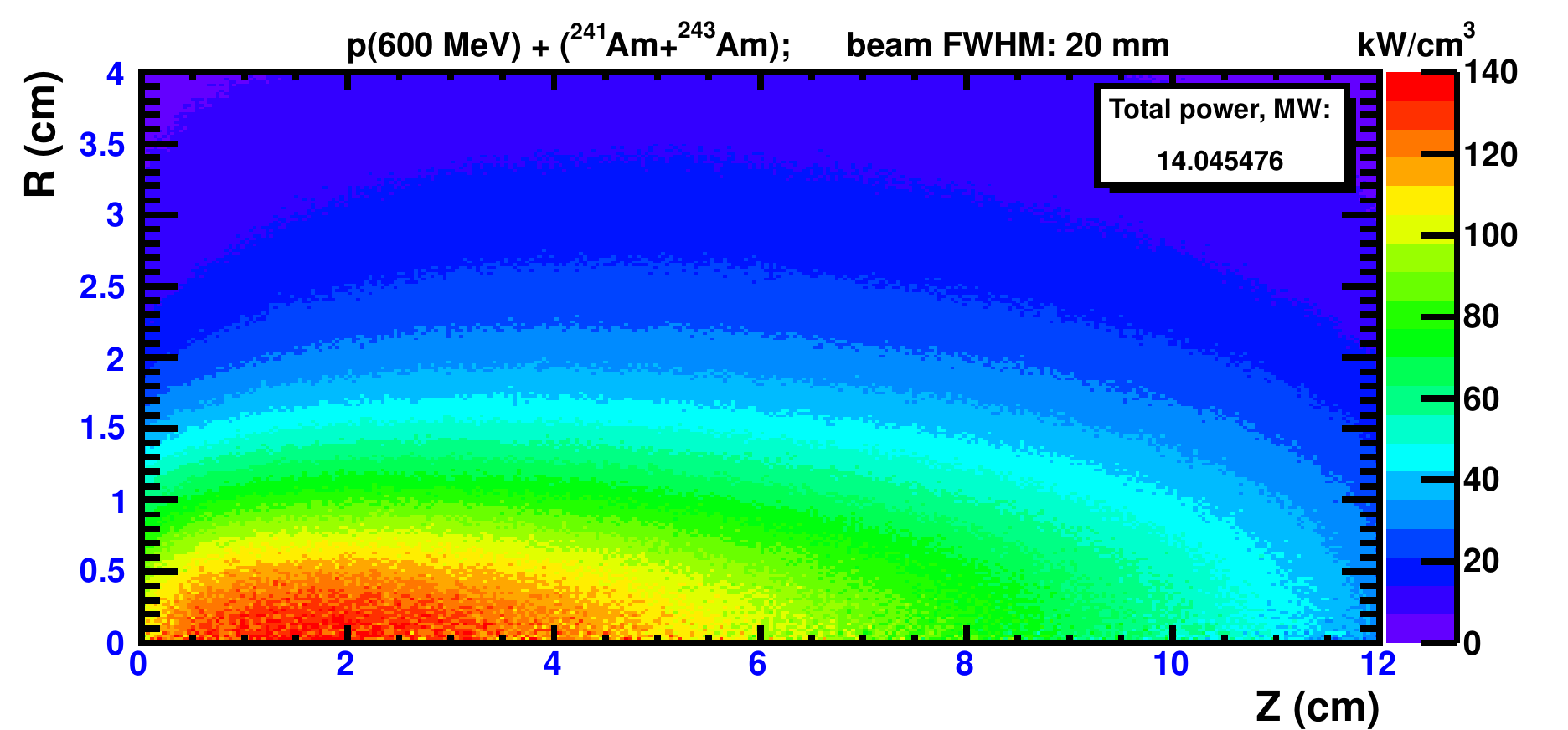}
\includegraphics[height=0.23\textheight]{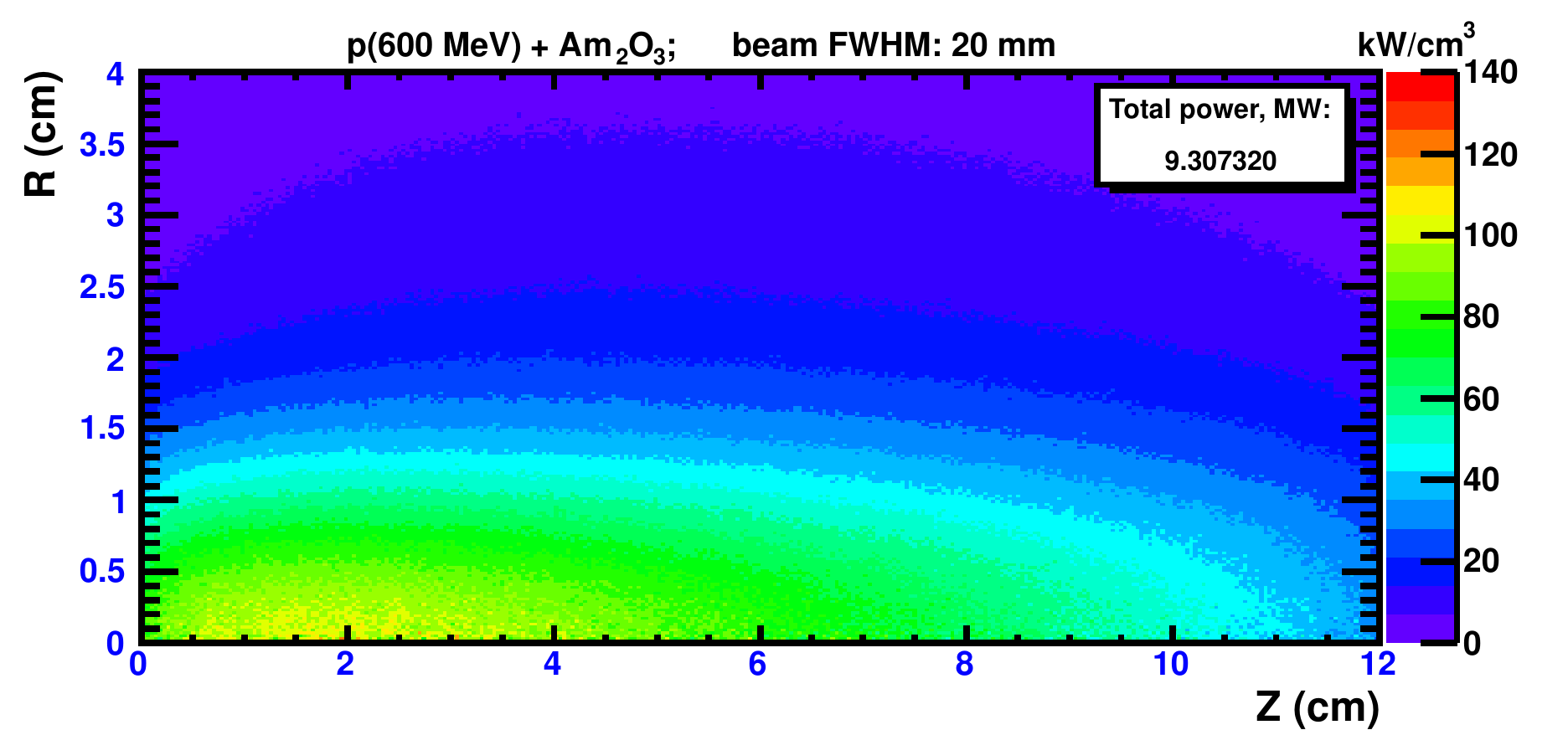}
\caption{Distribution of heat deposition inside cylindrical targets made of
$^{\rm nat}$U, $^{241}$Am, mixture of $^{241}$Am and $^{243}$Am and Am$_2$O$_3$, 
all of the same dimensions specified in Table~\ref{table:Am_vs_U}.
The targets are irradiated by 600~MeV proton beam with current of 10~mA.}
\label{fig:rod_eDep}
\end{centering}
\end{figure*}

Calculated spatial distributions of heat deposition inside the considered
targets are presented in Fig.~\ref{fig:rod_eDep}. As expected, in fissile spallation
targets a significant energy is deposited due to fission reactions~\cite{Malyshkin2012}. The
difference between fission cross section on Am and U leads to significantly
larger energy deposition in the americium targets (11.9--16.1~MW, depending on
the isotope composition) compared to the uranium target (7.7~MW). Therefore, 
designing a cooling system for the Am target may cause a serious problem.

The values of the average neutron flux ($1.32\cdot10^{16}$~n/s/cm$^2$) and heat
deposition (11.9~MW) calculated for a pure $^{243}$Am target are lower compared to a pure $^{241}$Am target.
This is explained by a lower fission cross section of $^{243}$Am compared to $^{241}$Am,
see Fig.~\ref{fig:ncapture_vs_fission}. The corresponding values for the target made of
the mixture of isotopes, $^{241}$Am+$^{243}$Am, are intermediate ($1.44\cdot10^{16}$~n/s/cm$^2$ and 14~MW)
with respect to the monoisotopic targets.

It is expected that an increased number of fission reactions in a spallation target makes possible to boost the
transmutation rate of americium contained in the target. However, due to the additional energy released in
fission events the heat deposition in the target rises too. As seen from Fig.~\ref{fig:rod_eDep},
the energy deposition in U and $^{241}$Am targets in the hottest region located close to the
target axis exceeds $100$~kW/cm$^3$, which looks very problematic from technical point of view. 
Therefore, more sophisticated target systems with a reduced energy deposition per fissioned Am nucleus are needed.

By this reason we extended our calculations for two U targets containing $^{241}$Am as proposed 
in~\cite{Malyshkin2012a} and a pure $^{243}$Am target mentioned above. Two targets containing $^{241}$Am  considered in~\cite{Malyshkin2012a} were schematically designed as following. In the first case $^{241}$Am was
uniformly mixed with U with 10\% mass concentration and in the second case a cylindrical core ($V = 200$ cm$^3$)
of $^{241}$Am is placed in the hottest region of the uranium target. In both
cases the target has the radius of 10~cm and length of 20~cm. The heat deposition values per
burned Am nucleus in these targets are several times higher than the
corresponding values for the pure americium targets because of
additional fission events of uranium nuclei which also produce neutrons.

The numbers of Am fission events per beam particle $N_{fis}$ are
listed in Table~\ref{table:Am_vs_U} for the six targets containing Am and
also for the $^{\rm nat}$U target taken as a reference case, where, accordingly, U fission events are counted.
As seen from the table, depending on isotope composition 2 to 3~times more fission events per beam proton are
estimated for the pure Am targets compared to the uranium target.
Calculated energy deposition per fissioned Am nucleus, $Q/N_{fis}$, is also
given in Table~\ref{table:Am_vs_U}. For the pure Am target this value (203~MeV) is
by 30\% lower than for the uranium one (279~MeV). As expected, much more energy is deposited per Am
fission event in two U+Am targets ($Q/N_{fis}=901$~MeV and 739~MeV), because only a part of fission events correspond to 
Am and most of them to U, according to their concentration in the targets. 

Since Am$_2$O$_3$ target contains less Am nuclei than the pure Am target, this
leads to a lower number of fission events and results, correspondingly, in a lower
neutron flux ($1.06\cdot10^{16}$~n/s/cm$^2$) and lower energy deposition (9.3~MW).
The heat deposition per fission event $Q/N_{fis}=229$~MeV calculated for
Am$_2$O$_3$ target is close to the $Q/N_{fis}$ value for the pure
$^{241}$Am+$^{243}$Am target.

Finally, burning rates of Am calculated for 10~mA proton beam are presented in
Table~\ref{table:Am_vs_U}. It was assumed that the burning rates are proportional
to the number of Am fission events $N_{fis}$  in a corresponding target. As found, more than 0.5~kg of
$^{241}$Am can be transmuted per month in the spallation target containing exclusively this isotope.
Lower burning rates are estimated for other considered geometry and material options, mostly
due to a lower Am content. The amount of americium transmuted during the first month of operation
$\rm{d}m/\rm{d}t(t=0)$ can be used to calculate the amount of Am burned during the first year.
In this estimation it is assumed that the amount of Am $m(t)$ decreases
exponentially, $m(t) = m_0 \exp(-t/\tau)$, from its initial amount $m_0$ with the
characteristic time $\tau = m_0/(\rm{d}m/\rm{d}t(t=0))$. The corresponding results are listed in
the last column of Table~\ref{table:Am_vs_U} which gives the minimum annual consumption
of Am in the considered spallation targets. Indeed, a possibility to upload additional quantities of Am to the target
(e.g. on the regular monthly basis) can be considered, thus substantially increasing the transmutation capability 
of the ADS system. One can conclude, that Am can be more efficiently burned in the 
spallation targets made of pure Am (more than 4~kg of Am per year) compared to mixed U+Am targets. 
However, the use of larger amounts of pure Am is restricted due to criticality issues.

\begin{table*}[!htb]
\caption{Initial amount of Am, number of fission reactions on Am per beam proton, $N_{fis}$, heat deposition per
fissioned Am nucleus, $Q/N_{fis}$, the burning rate of Am at the beginning of operation and the amount
of Am transmuted in the first year of operation for various target options.}
\centering
\begin{threeparttable}
\begin{tabular}{lccccccc}
\noalign{\smallskip}
\hline\noalign{\smallskip}
Material& Length &  Radius  & Initial   & $N_{fis}$     & $Q/N_{fis}$  & $\rm{d}m/\rm{d}t(t=0)$ & Burned in \\
        &  (cm)  &   (cm)   & Am mass   &               &              & Burning rate & first year      \\
        &        &          & (kg)      &               & (MeV)        & (g/month)    & (kg)            \\
\hline\noalign{\smallskip}
$^{\rm nat}$U & 12  & 4    & 0            & 2.74\tnote{*} & 279\tnote{*} & ---       & ---          \\
\noalign{\smallskip}
U+$^{241}$Am (10\%) & 20 & 10 & 11.7          & 1.84          & 901          & 120        & 1.4     \\
U+$^{241}$Am (core) & 20 & 10  & 2.68          & 2.25          & 739          & 147        & 1.6      \\
\noalign{\smallskip}
Pure $^{241}$Am  & 12  & 4        & 8.24          & 7.91          & 203          & 514        & 4.3       \\
Pure $^{243}$Am  & 12  & 4        & 8.24          & 5.49          & 216          & 357        & 3.4       \\
$^{241}$Am$ + ^{243}$Am & 12  & 4  & 8.24          & 6.74          & 208          & 438        & 3.9      \\
Am$_2$O$_3$ & 12  & 4   & 6.46          & 4.07          & 229          & 265        & 2.5             \\
\hline\noalign{\smallskip}
\end{tabular}
\begin{tablenotes}
\item[*] Number of fission reactions on U nuclei per beam proton. For all other target options
$N_{fis}$ corresponds exclusively to Am fission events.
\end{tablenotes}
\end{threeparttable}
\label{table:Am_vs_U}
\end{table*}

As shown in our previous publication~\cite{Malyshkin2012a}, in the spallation 
targets made of $^{\rm nat}$U the neutron production is significantly enhanced 
with respect to the tungsten targets of the same size due to additional 
contribution of neutron-induced fission of uranium nuclei. In the present study 
we have found that even more neutrons are produced by fission reactions in 
targets made of pure $^{241}$Am and $^{243}$Am. This follows from 
Table~\ref{table:n_prod_and_abs}, where the numbers of neutrons produced or 
absorbed in various nuclear reactions are given per beam particle for 
$^{\rm nat}$U, $^{241}$Am and $^{243}$Am targets of the same size. This means 
that even small Am targets are highly efficient in incinerating Am in fission 
reactions. The number of neutrons which escape from the targets are also 
presented in Table~\ref{table:n_prod_and_abs}. Obviously, the number 
of leaking neutrons is equal to the difference between the number of produced 
and absorbed neutrons. The calculations show that about 45\% more neutrons are 
produced in $^{241}$Am target than in $^{\rm nat}$U one. This means that 
$^{241}$Am target can serve as an intensive neutron source which can be used 
for various applications.

When dealing with spallation targets made of fissile materials one should 
consider also an additional heat produced in fission reactions. The total heat 
deposition $Q$ and heat per leaked neutron are presented in 
Table~\ref{table:average_numbers} for the same targets as before. One can see
that about twice as much heat is generated in the $^{241}$Am target as compared 
with $^{\rm nat}$U one. This means that a sophisticated cooling system is 
required for its operation.

\begin{table}[htb]
\caption{Average contributions to neutron production and absorption from different reaction 
channels and the number of leaking neutrons for $^{\rm nat}$U, $^{241}$Am and $^{243}$Am
targets with the length of 12~cm and radius of 4~cm. All numbers are given per beam proton.}
\centering
\begin{threeparttable}
\begin{tabular}{lccc}
\noalign{\smallskip}
\hline\noalign{\smallskip}
	              &  $^{\rm nat}$U &  $^{241}$Am &  $^{243}$Am	\\
\hline\noalign{\smallskip}
p + A 	              &  11.77     &  9.01	 &  9.34	\\
(n,2n) 	              &  0.44	   &  0.04	 &  0.11	\\
(n,3n) 	              &  0.15	   &  0.01	 &  0.02	\\
(n,4n) 	              &  0.09	   &  0.06	 &  0.06	\\
(n,$>$4n)             &  3.80	   &  1.67	 &  1.91	\\
(n,fission)\tnote{*} &  4.17	   &  19.54 	 &  13.79 	\\
produced by other     &  0.15      &   0.26      &   0.17       \\
particles             &            &             &              \\
(n,$\gamma$)         &  -0.45	   &  -1.45	 &  -1.00	\\
leak 	              &  20.12	   &  29.14	 &  24.40	\\
\hline\noalign{\smallskip}
\end{tabular}
\begin{tablenotes}
\item[*] Only from neutron-induced fission below 20~MeV.
\end{tablenotes}
\end{threeparttable}
\label{table:n_prod_and_abs}
\end{table}

\begin{table}[htb]
\caption{Average heat deposition $Q$ and heat deposition per leaked neutron 
$Q/N$ calculated with MCADS for $^{\rm nat}$U,   $^{241}$Am  and  $^{243}$Am 
targets with the length of 12~cm and radius of 4~cm. All values are given per 
beam proton.}
\centering
\begin{tabular}{lccc}
\noalign{\smallskip}
\hline\noalign{\smallskip}
            &  $^{\rm nat}$U  &  $^{241}$Am  &  $^{243}$Am  \\
\hline\noalign{\smallskip}
$Q$ (MeV)   &  766        &  1608        &  1185        \\
$Q/N$ (MeV) &  38.1       &  55.2        &  48.6        \\
\hline
\end{tabular}
\label{table:average_numbers}
\end{table}

\section{Spallation targets made of Am with U booster and Be reflector}\label{boost+refl}

\begin{figure}[!htb]
\begin{centering}
\includegraphics[width=1.0\columnwidth]{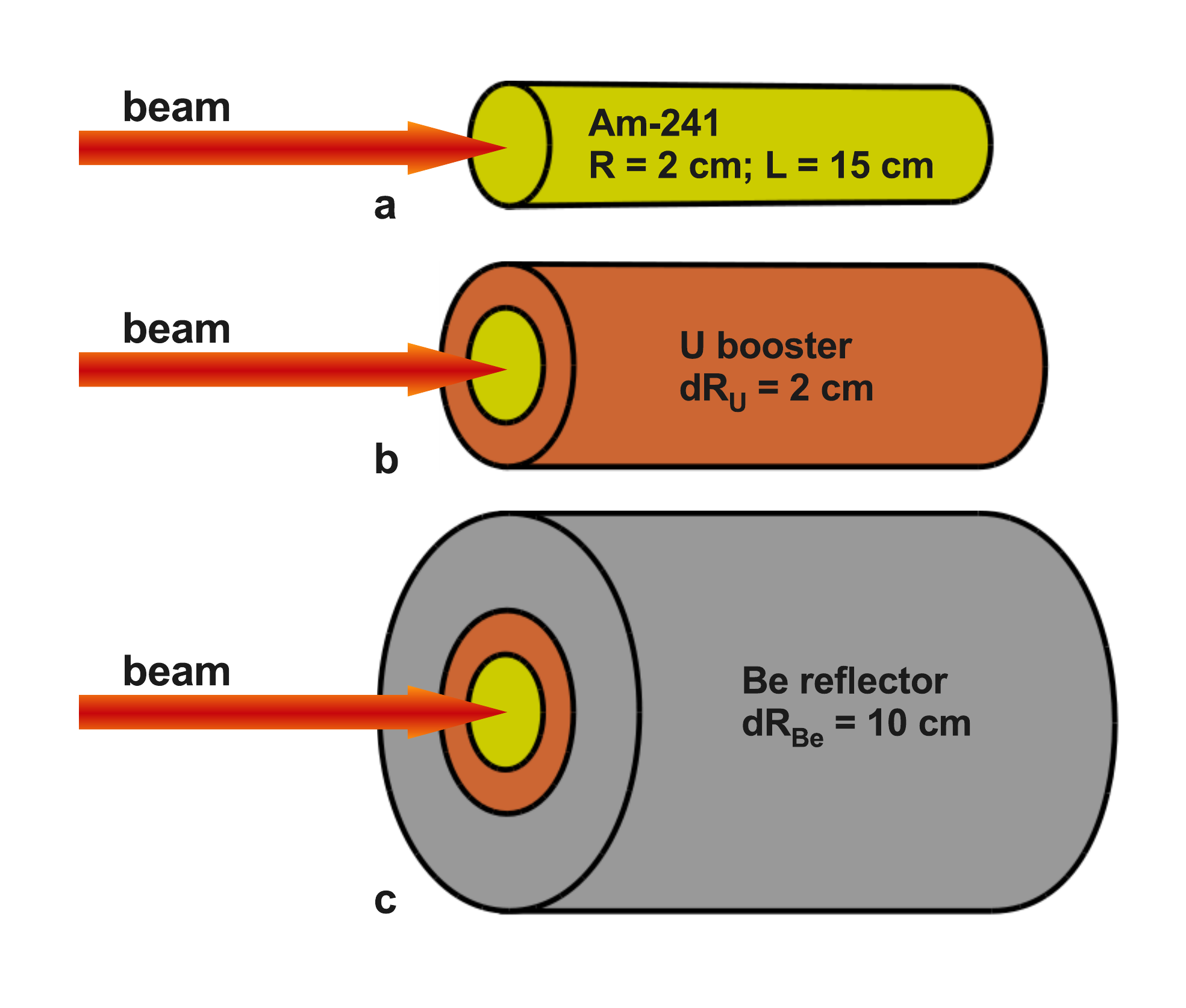}
\caption{Geometry options: bare $^{241}$Am cylinder (option {\it a}),
$^{241}$Am cylinder covered by a U booster (option {\it b}) and
$^{241}$Am cylinder covered by a $^{\rm nat}$U booster and a $^{9}$Be reflector
(option {\it c}).
The arrows indicate direction of the proton beam.}
\label{fig:boost_refl}
\end{centering}
\end{figure}

Several advanced geometry options of the spallation target can be considered to increase the burning rate of Am by 
enhancing the neutron flux inside the target. Two such options are considered below.
The first target option ({\it b}) consists of an $^{241}$Am cylinder covered by a 2~cm thick $^{\rm nat}$U booster. 
The second option ({\it c}) is additionally covered by a $^9$Be
reflector which is 10~cm thick. The core of the both target options consist of a $^{241}$Am cylinder (rod) with the length of
150~mm and radius of 20~mm, which is also used alone as a reference target option ({\it a}) for comparison. 
All three target options are schematically shown in Fig.~\ref{fig:boost_refl} and their parameters are listed in
Table~\ref{table:advanced}. 

\begin{table*}[!htb]
\caption{Initial amount of Am, number of fission events on Am per beam proton, $N_{fis}$, heat deposition per
fissioned Am nucleus, $Q/N_{fis}$, the burning rate of Am at the beginning of operation and the amount
of Am transmuted in the first year of operation for various target options.}
\centering
\begin{threeparttable}
\begin{tabular}{lccccc}
\noalign{\smallskip}
\hline\noalign{\smallskip}
Material                         & Initial       & $N_{fis}$     & $Q/N_{fis}$  & $\rm{d}m/\rm{d}t(t=0)$ & Burned in \\
                                 & Am mass       &               &              & Burning rate & first year      \\
                                 & (kg)          &               & (MeV)        & (g/month)    & (kg)            \\
\hline\noalign{\smallskip}
$^{241}$Am ({\it a})             &    2.58       &    3.00       &    241       &     194      &      1.53      \\
$^{241}$Am + U booster ({\it b}) &    2.58       &    3.79       &    229       &     245      &      1.76       \\
$^{241}$Am + U booster           &    2.58       &    3.95       &    228       &     256      &      1.79       \\
+ Be reflector ({\it c})         &               &               &              &              &                 \\
\hline\noalign{\smallskip}
\end{tabular}
\end{threeparttable}
\label{table:advanced}
\end{table*}

Initially all the three targets contain four times less $^{241}$Am  (2.58~kg) compared
to the pure $^{241}$Am target considered in Sec.~\ref{am:simple}. However, the number of fission events per beam proton
$N_{fis}$ in the target (c) is only twice as low as in the target of Sec.~\ref{am:simple}. This means that Am
is burned more efficiently in the presence of the booster, or with both booster and reflector.
At the same time $Q/N_{fis}$ calculated in the Am core for option (c) is comparable to the same parameter for the
pure $^{241}$Am target considered in Sec.~\ref{am:simple}. The total power deposition in Am volume calculated
for options (b) and (c)  (8.67 and 8.98 MW, respectively) is much less than for the pure $^{241}$Am target (16.08 MW).
This indicates the advantages of more complicated target geometries (b) and (c) with respect to simple geometry,
since additional neutrons are produced in the U booster. We should also note that no essential improvement in the target 
performance was achieved by the addition of the Be reflector. 

The absolute Am burning rates are smaller ($\sim200$~g/month) than for the simple target ($\sim500$~g/month). However,
the relative burning rates are higher for the advanced options. Indeed, about 42\% of the initial Am mass
is burned in the first year of ADS operation in the simple pure $^{241}$Am target, as compared with 69\% in the advanced
target (c).

\begin{figure*}[!htb]
\begin{centering}
\includegraphics[height=0.31\textheight]{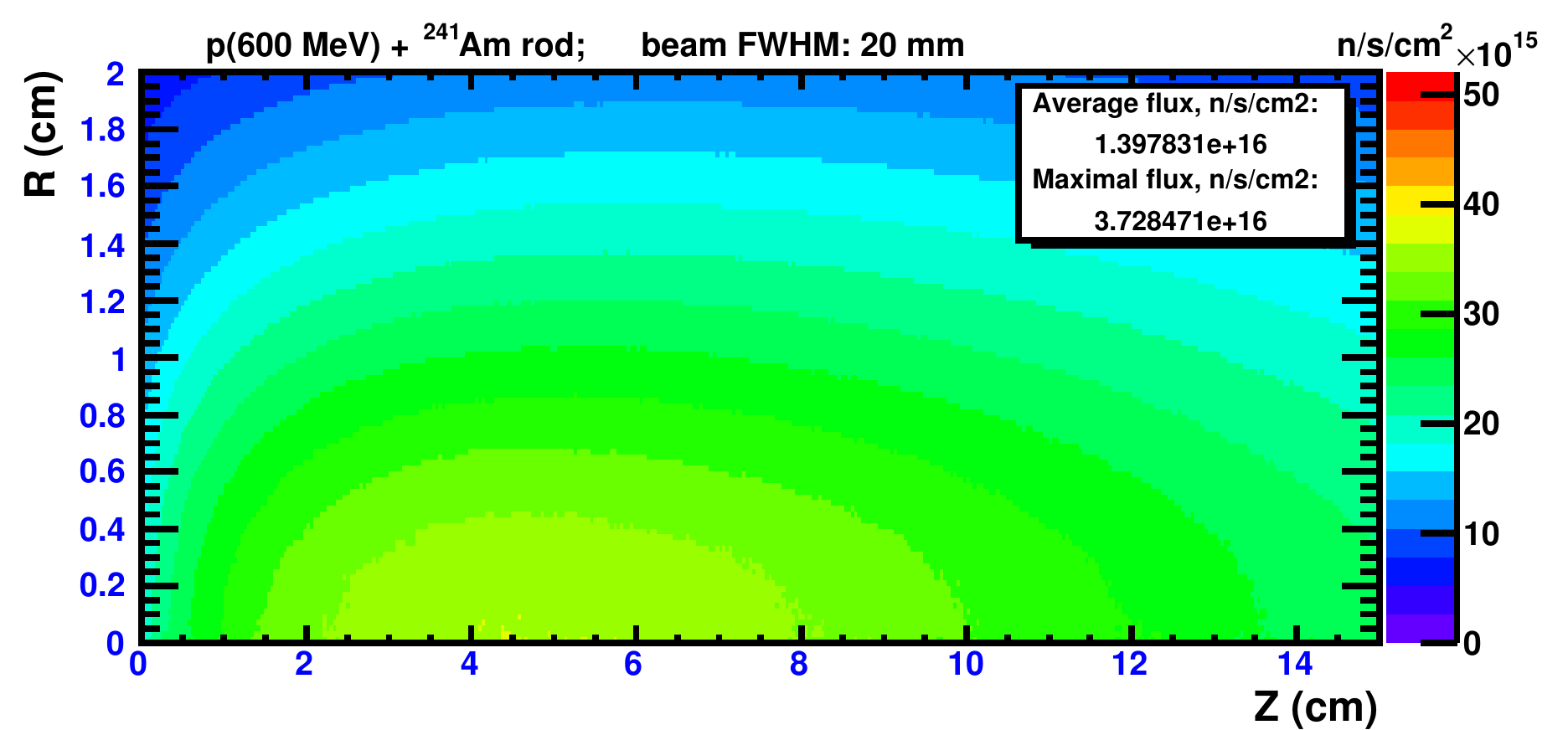}
\includegraphics[height=0.31\textheight]{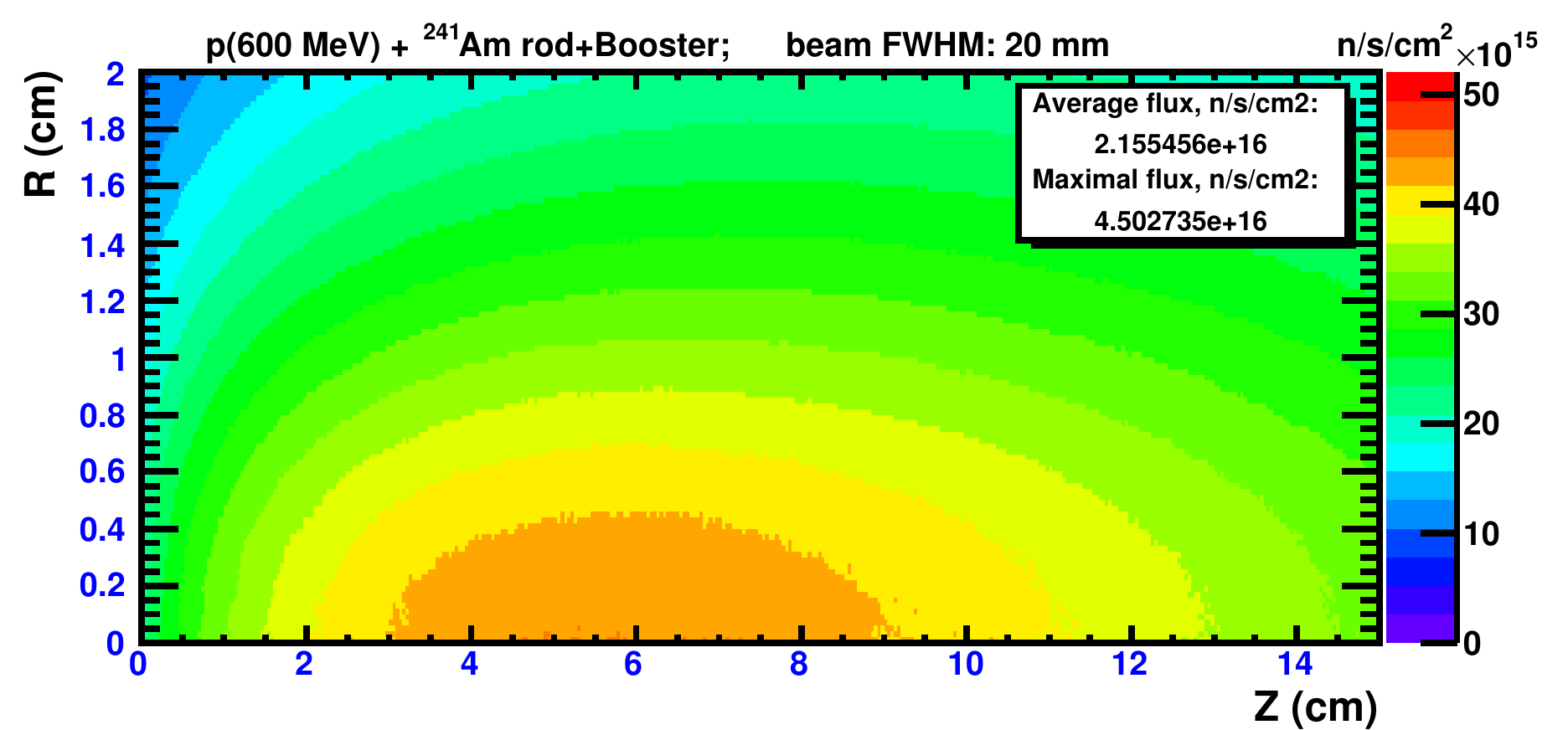}
\includegraphics[height=0.31\textheight]{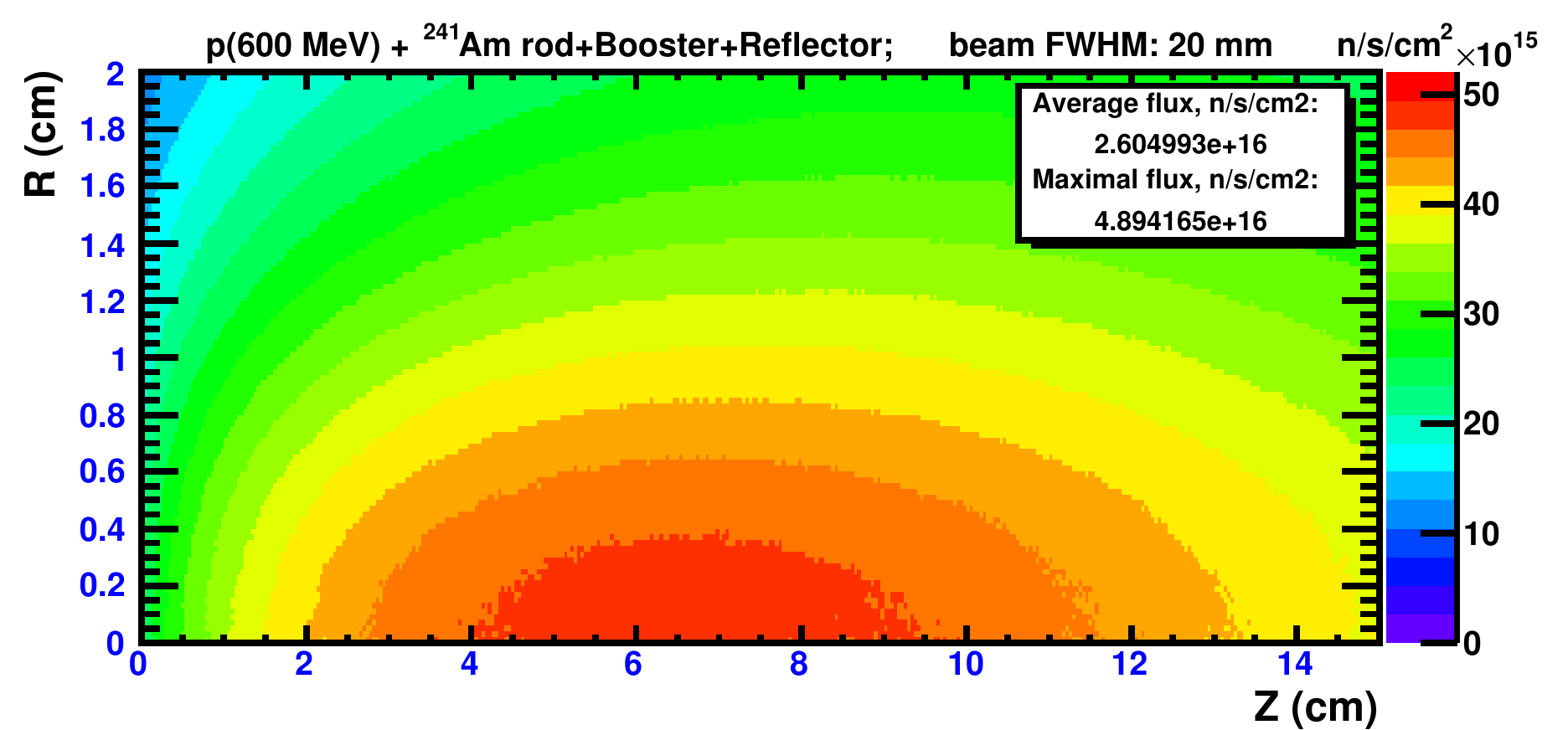}
\caption{Distribution of neutron flux inside $^{241}$Am cylindrical target (option {\it a}),
$^{241}$Am cylinder covered by a $^{\rm nat}$U booster (option {\it b}) and $^{241}$Am cylinder covered
by the $^{\rm nat}$U booster and a $^9$Be reflector (option {\it c}).
The distributions are calculated for the targets irradiated by 600~MeV proton beam of 10~mA.}
\label{fig:boost_refl_nCF}
\end{centering}
\end{figure*}

\begin{figure*}[!htb]
\begin{centering}
\includegraphics[height=0.31\textheight]{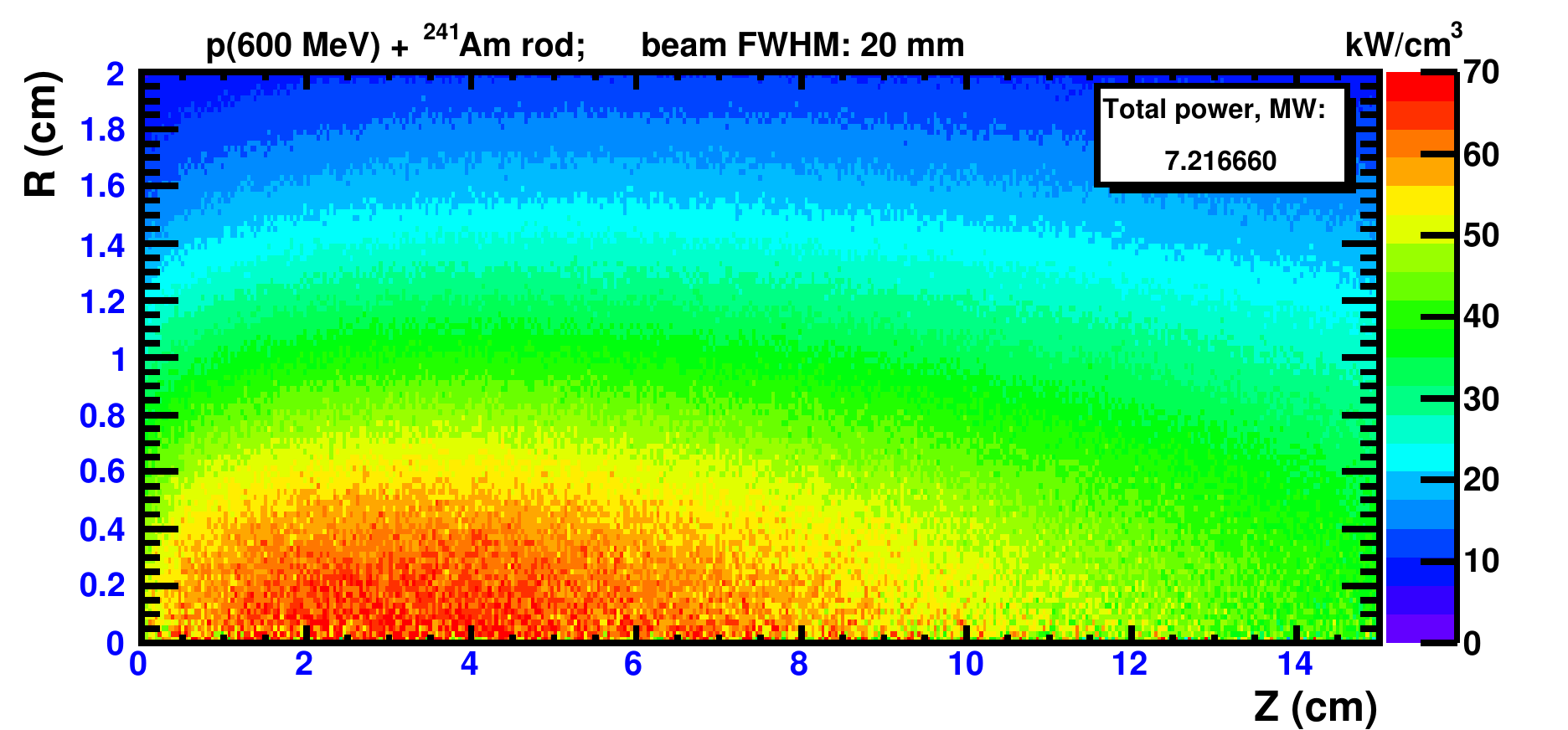}
\includegraphics[height=0.31\textheight]{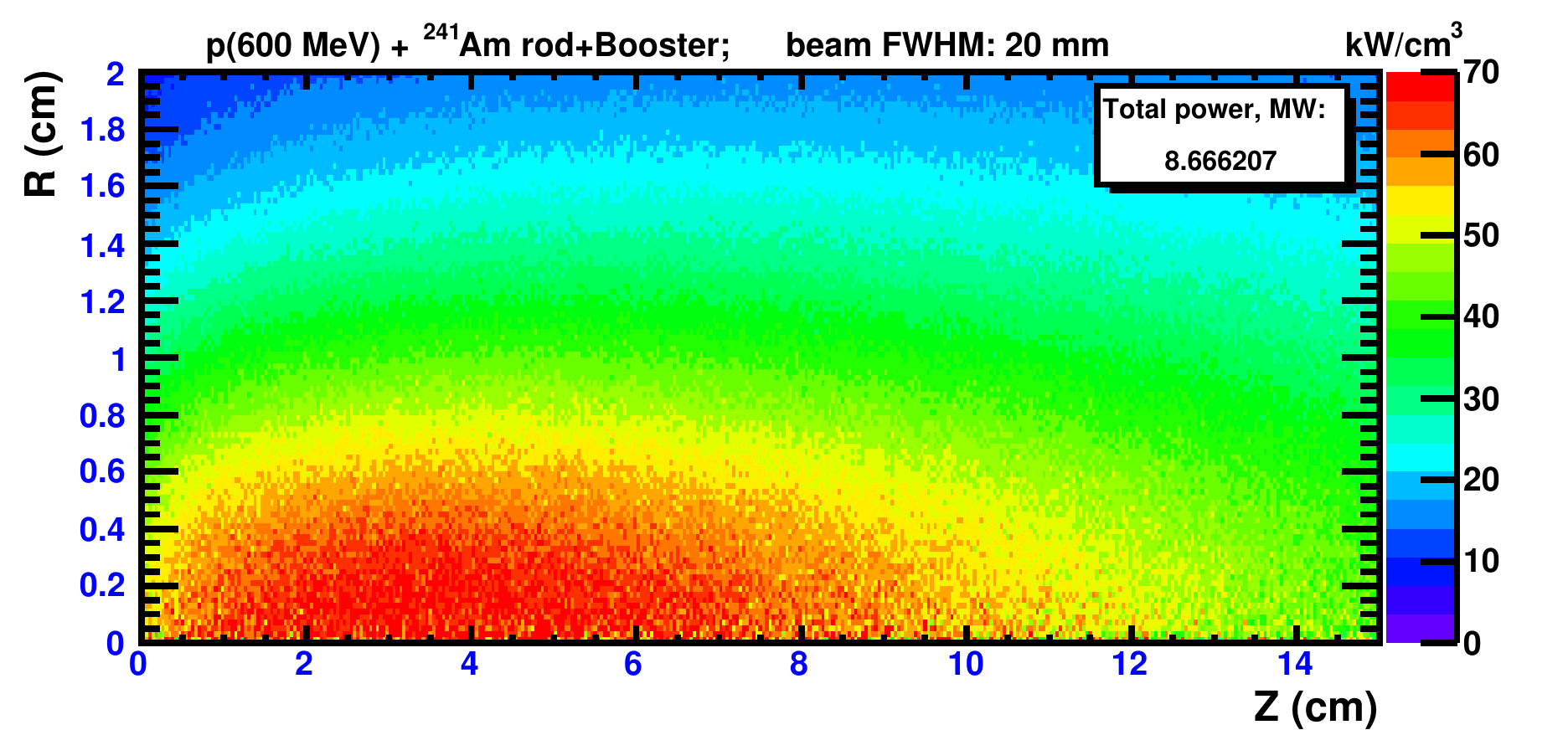}
\includegraphics[height=0.31\textheight]{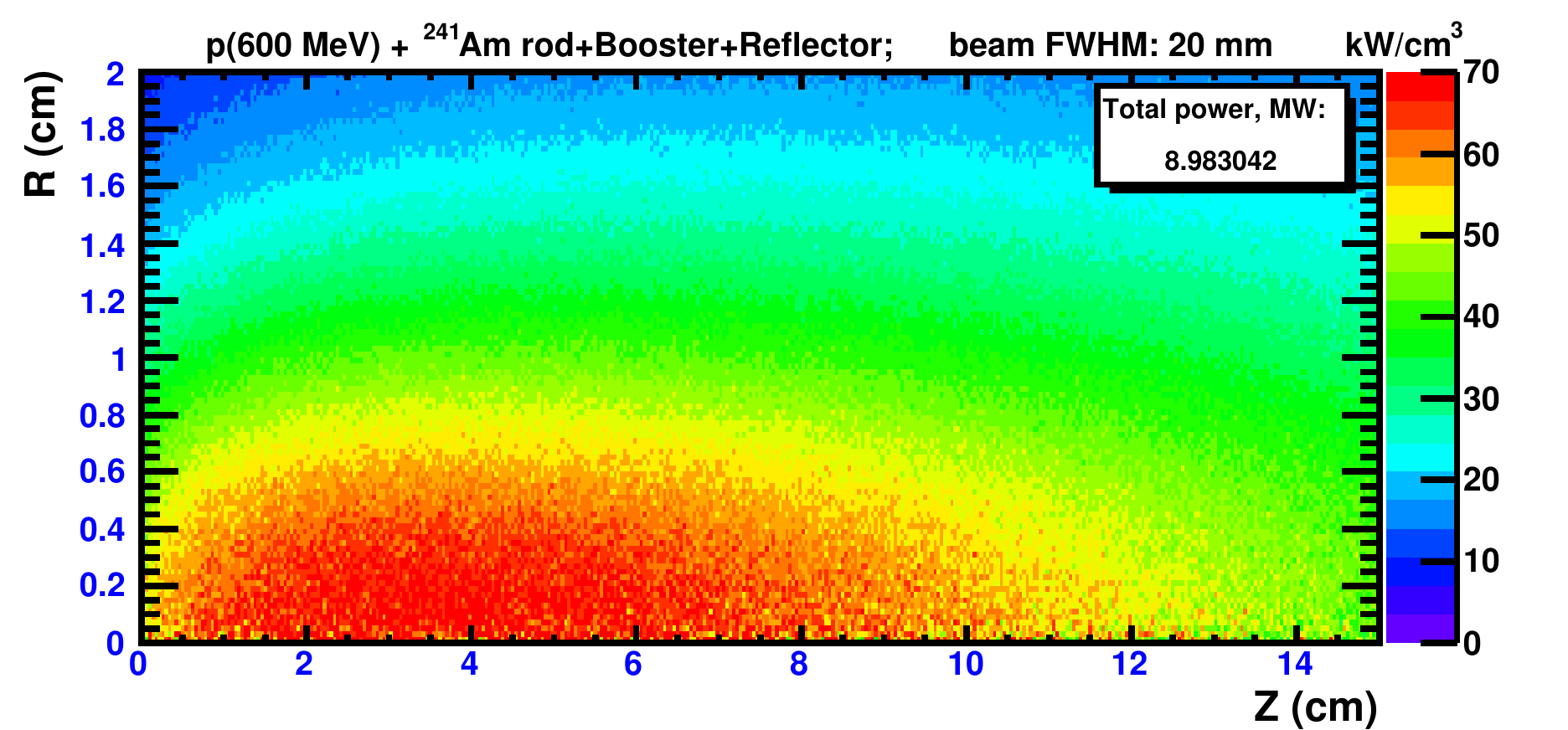}
\caption{Distribution of heat deposition inside $^{241}$Am cylindrical target (option {\it a}),
$^{241}$Am cylinder covered by a $^{\rm nat}$U booster (option {\it b}) and $^{241}$Am cylinder covered
by the $^{\rm nat}$U booster and a $^9$Be reflector (option {\it c}).
The distributions are calculated for the targets irradiated by 600~MeV proton beam of 10~mA.}
\label{fig:boost_refl_eDep}
\end{centering}
\end{figure*}

The distributions of the neutron flux and energy deposition for the options (a)--(c) are shown in
Figs.~\ref{fig:boost_refl_nCF} and ~\ref{fig:boost_refl_eDep}. As seen in these figures, the distributions are more
uniform both along the axis of the target and its radius. One can see that adding the booster and the reflector leads
to increased average neutron flux by about 50\% and 90\%, respectively. The highest neutron flux of 
$4.9\cdot10^{16}$~n/s/cm$^2$ is reached locally in the target (c) with its average value of 
$2.6\cdot10^{16}$~n/s/cm$^2$.  The highest volumetric energy deposition 
is below 70~kW/cm$^3$ for the options (b)--(c), which is twice as low as the value of 140~kW/cm$^3$ 
calculated for the simple $^{241}$Am target.

\section{Conclusions}
\label{sec:conclusion}

In this paper we have applied the MCADS model based on the Geant4 toolkit for calculating neutron fields and
heat deposition in spallation targets containing uranium and americium. We have investigated the criticality of
such targets using the Monte Carlo method. We have demonstrated that the critical mass of extended 
spallation targets containing $^{241}$Am can be evaluated with MCADS by calculating 
the corresponding neutron multiplication factor as a function of time elapsed since
the impact of a beam proton. The MCADS results for the criticality of targets
made of $^{241}$Am provide a guide-line for further studies for their optimization and safe operation in 
a deep subcritical mode.

Depending on the composition of the target material (pure $^{241}$Am, 
$^{243}$Am, their mixture or Am$_2$O$_3$), 2.5--4.3~kg of Am can be transmuted 
into short-lived or stable fission fragments per 
year of operation in the spallation target of the ADS facility irradiated by 
600~MeV proton beam of 10~mA. The results of simulations with 
targets of different material composition show that the highest rate of 
americium incineration is achieved in the targets made of pure americium. 
As demonstrated by simulations, when Am cylinder is covered by the U booster 
and shielded by the Be reflector, the relative annual incineration rate of Am 
increases up to 69\%. The burning rate may be increased by $\sim 50$\% if 
uploading of additional quantities of Am to the spallation target is made on a 
regular basis. Higher incineration rates can be obtained by increasing the 
neutron multiplication factor up to $k\sim 0.8$. However, a high energy 
deposition in these targets creates serious challenges to their cooling systems.
Additional studies of technological issues related to the high heat deposition 
and also to the radiation damage of target materials due to a very high neutron 
flux predicted in the target are needed to prove the viability of the proposed 
concept.

In the uranium targets with small admixture of Am long-lived isotope $^{237}$Np 
is produced following the capture of two neutrons by $^{235}$U and the 
subsequent beta-decay. However, our estimations show that the amount of 
$^{237}$Np produced after a year of irradiation does not exceed 100~grams, 
which is much less than the mass of transmuted MA. The neutron capture on Am 
nuclei leading to the production of heavier long-lived MA does not change the 
total amount of  MA in the target.

In this paper we did not discuss a well-known solution where a spallation target of a full-scale ADS facility 
is surrounded by an extended subcritical reactor core~\cite{Tsujimoto2004,Ismailov2011}. In this case one could
not only use neutrons escaped from the spallation target, but also use additional neutrons produced in the 
reactor core to burn MA placed there. This may significantly, by a factor of 10 or more, increase the 
amount of burned MA~\cite{Tsujimoto2004} as compared with burning only in the spallation target. 
Several such ADS facilities can solve the problem of utilization of MA produced in thermal reactors. 
The thermal energy produced in the spallation target and reactor core can be converted to electricity
in order to cover (at least partially) the energy consumed by the accelerator.

Finally, one can note that the highest neutron flux ($4.9\cdot10^{16}$~n/s/cm$^2$) is reached in the target with the
booster and reflector, and the average neutron flux ($2.6\cdot10^{16}$~n/s/cm$^2$) is also high
in this target. Therefore, the ADS facilities can be used to study the properties of materials under the 
impact of intense irradiation by fast neutrons, as well as for basic research.

\section*{Acknowledgments}

Our calculations were performed at the Center for Scientific Computing (CSC) of
the Goethe University, Frankfurt am Main. We are grateful to the staff of
the Center for support.
We thank to E.~Mendoza and D.~Cano-Ott (CIEMAT, Madrid) for providing the
evaluated neutron data in Geant4-format. 
We are also thankful to Siemens AG for financial support.

\section*{References}
\bibliography{ADS_Literature_v2,Americium_v2}

\end{document}